\documentclass[3p,times,twocolumn]{elsarticle}

\usepackage{ecrc}
\usepackage{hyperref}

\volume{N}

\firstpage{1}

\journalname{Nuclear Physics B Proceedings Supplement}

\runauth{Miguel Nebot}

\jid{nuphbp}

\jnltitlelogo{Nuclear Physics B Proceedings Supplement}

\usepackage{xcolor}
\usepackage{mathrsfs}
\usepackage{amsmath}
\usepackage{multirow}
\usepackage{graphicx}

\usepackage{amssymb}
\usepackage{amsthm}

\usepackage[figuresright]{rotating}

\newcommand{\CKM}{V}

\newcommand{\N}{N}
\newcommand{\Nu}{\N_{u}}
\newcommand{\Nd}{\N_{d}}
\newcommand{\M}{M}
\newcommand{\Mu}{\M_{u}}
\newcommand{\Md}{\M_{d}}
\newcommand{\Nud}{\N_{u}^\dagger}
\newcommand{\Ndd}{\N_{d}^\dagger}
\newcommand{\gR}{\gamma_R}
\newcommand{\gL}{\gamma_L}
\newcommand{\QLo}{Q_{L}^{0}}
\newcommand{\QLbo}{\overline{Q_{L}^{0}}}
\newcommand{\uLbo}{\overline{u_{L}^{0}}}
\newcommand{\uRbo}{\overline{u_{R}^{0}}}
\newcommand{\dLbo}{\overline{d_{L}^{0}}}

\newcommand{\uRo}{u_{R}^{0}}

\newcommand{\dRo}{d_{R}^{0}}
\newcommand{\dLo}{d_{L}^{0}}
\newcommand{\LLo}{L_{L}^{0}}
\newcommand{\LLbo}{\overline{L_{L}^{0}}}

\newcommand{\nRo}{\nu_{R}^{0}}
\newcommand{\lRo}{\ell_{R}^{0}}

\newcommand{\nchck}{{\checkmark}}
\newcommand{\grchck}{{\footnotesize\color{gray}\nchck}}

\newcommand{\refeq}[1]{eq.(\ref{#1})}

\graphicspath{{./Figs/}}

\begin{document}

\begin{frontmatter}



\dochead{}

\title{Constraints on a Class of Two-Higgs Doublet Models with tree level FCNC}


\author{Miguel Nebot}

\address{Centro de F\'isica Te\'orica de Part\'iculas,\\
Instituto Superior T\'ecnico, Universidade de Lisboa,\\
 Av. Rovisco Pais, P-1049-001 Lisboa,
Portugal}

\begin{abstract}
We analyse a class of two Higgs doublet models where flavour-changing neutral currents (FCNC) are present at tree level in a mixing-suppressed manner. In this class of models, because of a discrete symmetry imposed on the lagrangian, the FCNC couplings in the quark and lepton sector are fixed in terms of the corresponding mixing matrix (CKM or PMNS), the fermion masses and the ratio $v_2/v_1$ of the vacuum expectation values of the neutral scalars. A large number of processes, including tree and loop level transitions mediated by the new charged or neutral scalars are used as constraints. It is shown that among the interesting phenomenological prospects for these models, the new scalars may have masses within experimental reach.
\end{abstract}

\begin{keyword}
Two Higgs Doublet Models \sep Symmetries \sep FCNC \sep Minimal Flavour Violation

\end{keyword}

\end{frontmatter}


\section{Introduction}
\label{sec:Introduction}
The discovery of a particle with mass around 125 GeV and properties consistently compatible with those of the Standard Model (SM) Higgs boson by the ATLAS and CMS collaborations \cite{Aad:2012tfa,Chatrchyan:2012tx} has brought much attention to studies of the scalar sector beyond the SM. Scalar sectors more complex than the SM one are present in many of its extensions. Two Higgs doublet models (2HDM) are a particularly interesting case \cite{Lee:1973iz,Branco:2011iw}; the general 2HDM without additional symmetries introduces tree level flavour-changing neutral currents (FCNC): if they are unsuppressed, conflict with experimental results in the flavour sector is to be expected \cite{Crivellin:2013wna}. Direct avoidance of tree level FCNC in the 2HDM can be achieved through the imposition of a discrete $\mathbb Z_2$ symmetry, producing ``Natural Flavour Conservation'' \cite{Glashow:1976nt,Paschos:1976ay}; another possiblity is the assumption of flavour alignment of the Yukawa couplings \cite{Pich:2009sp} (however, this assumption, in its simplest form, is not well behaved under renormalization group evolution \cite{Ferreira:2010xe}). An alternative to complete removal of tree level FCNC is the possibility of having them in a controlled manner \cite{Joshipura:1990pi,Lavoura:1994ty,Branco:1996bq}. A particularly interesting case on which this contribution focusses is the class of so-called ``BGL models'', proposed by Branco, Grimus and Lavoura in \cite{Branco:1996bq} and further studied in \cite{Botella:2009pq,Botella:2011ne,Botella:2012ab,Botella:2014ska,Bhattacharyya:2014nja}, where the introduction of a symmetry yields tree level FCNC that depend on fermion mixings, fermion masses and the ratio of vacuum expectation values $v_2/v_1=\tan\beta$. Furthermore, a consistent renormalization group evolution treatment requires addressing both quark and lepton sectors \cite{Botella:2011ne}. In the following a concise introduction to BGL models is presented, together with the abundant experimental constraints mainly coming from flavour physics that have to be considered before addressing the central scope of this work: a global study of the phenomenological prospects for the different models in this class.
\section{BGL models}
\label{sec:BGL}
Yukawa couplings in the general 2HDM have the following form:
\begin{multline}
\mathscr L_{\rm Y}=
-\QLbo\big(\Delta_1\tilde\Phi_1+\Delta_2\tilde\Phi_2\big)\uRo-\QLbo \big(\Gamma_1\Phi_1+\Gamma_2\Phi_2\big)\dRo\\
-\LLbo\big(\Sigma_1\tilde\Phi_1+\Sigma_2\tilde\Phi_2\big)\nRo-\LLbo \big(\Pi_1\Phi_1+\Pi_2\Phi_2\big)\lRo+\text{h.c.},
\label{eq:Yukawa:01}
\end{multline}
where\footnote{Generation indices are not displayed, fermion fields in \refeq{eq:Yukawa:01} are to be understood as 3-vectors; $\Delta_i$, $\Gamma_i$, $\Sigma_i$ and $\Pi_i$ are accordingly $3\times 3$ matrices.} $\QLo$ and $\LLo$ are the usual left-handed quark and lepton ${\rm SU}(2)_L$ doublets while $\uRo$, $\dRo$, $\nRo$ and $\lRo$ are the right-handed ${\rm SU}(2)_L$ singlets; $\Phi_1$ and $\Phi_2$ are the scalar ${\rm SU}(2)_L$ doublets and $\tilde\Phi_j=i\sigma_2{\Phi_j}^\ast$. Spontaneous electroweak symmetry breaking by the vacuum expectation values $\langle\mathbf{0}|\Phi_i|\mathbf{0}\rangle=\frac{e^{i\alpha_i}}{\sqrt 2}\left(\begin{smallmatrix}0\\ v_i\end{smallmatrix}\right)$ leaves a spectrum in the scalar sector consisting of one charged field $H^\pm$ and three neutral fields $H^0$, $R$ and $I$; we assume in addition that $H^0$ behaves as the SM Higgs, having in particular couplings to fermions proportional to their masses and $v=({v_1^2+v_2^2})^{1/2}\simeq 246$ GeV. For conciseness we focus in the quark sector: 
\begin{multline}
\mathscr L_{\rm Y}\supset
-\uLbo\frac{1}{v}\big(\Mu^0(v+H^0)+\Nu^0 R+i\Nu^0 I\big)\uRo\\
-\dLbo\frac{1}{v}\big(\Md^0(v+H^0)+\Nd^0 R+i\Nd^0 I\big)\dRo\\
-\frac{\sqrt 2}{v}\big(\uLbo\Nd^0\dRo-\uRbo{\Nu^0}^\dagger\dLo\big)H^++\text{h.c.}
\end{multline}
where\footnote{The difference of the phases of the vacuum expectation values is $\theta\equiv\alpha_1-\alpha_2$.}
\begin{equation}
\Mu^0=\frac{1}{\sqrt 2}\big(v_1\Delta_1+v_2e^{i\theta}\Delta_2\big),\ 
\Md^0=\frac{1}{\sqrt 2}\big(v_1\Gamma_1+v_2e^{i\theta}\Gamma_2\big),\label{eq:MassMat:01}
\end{equation}
and
\begin{equation}
\Nu^0=\frac{1}{\sqrt 2}\big(v_2\Delta_1-v_1e^{i\theta}\Delta_2\big),\
\Nd^0=\frac{1}{\sqrt 2}\big(v_2\Gamma_1-v_1e^{i\theta}\Gamma_2\big).\label{eq:NMat:01}
\end{equation}
Diagonalisation proceeds as usual
\begin{multline*}
U^\dagger_{uL}\Mu^0 U_{uR} \equiv \Mu= \text{diag}(m_u, m_c, m_t),\\ U^\dagger_{dL}\Md^0 U_{dR} \equiv \Md = \text{diag}(m_d, m_s, m_b),
\end{multline*}
where $\CKM\equiv U^\dagger_{uL}U_{dL}$ is the CKM mixing matrix. $\Nu^0$ and $\Nd^0$ are not, a priori, diagonalised:
\begin{equation*}
U^\dagger_{uL}\Nu^0 U_{uR}\equiv\Nu,\quad U^\dagger_{dL}\Nd^0 U_{dR}\equiv\Nd\,.
\end{equation*}
In terms of physical fields, we are thus left with
\begin{multline}
\mathscr L_{\rm Y}\supset
-\frac{1}{v}H^0 \big(\bar u \Mu u+\bar d \Md d\big)\\
-\frac{1}{v}R\bigg[\bar u\big(\Nu\gR+\Nud\gL\big)u+\bar d\big(\Nd\gR+\Ndd\gL\big)d\bigg]\\
+\frac{i}{v}I\bigg[\bar u\big(\Nu\gR-\Nud\gL\big)u-\bar d\big(\Nd\gR-\Ndd\gL\big)d\bigg]\\
-\frac{\sqrt 2}{v} H^+ \bar u \big( \CKM \Nd\gR-\Nud \CKM \gL\big)d+\text{h.c.}
\end{multline}
where tree level FCNC involving $R$ and $I$ are explicitly controlled by $\Nu$ and $\Nd$. Following the BGL proposal \cite{Branco:1996bq}, we impose symmetry under the following transformation:
\begin{equation}
Q_{Lj}^{0}\mapsto e^{i\tau}Q_{Lj}^{0}\,,\
u_{Rj}^{0}\mapsto e^{i2\tau}u_{Rj}^{0}\,,\ 
\Phi_{2}\mapsto e^{i\tau}\Phi_{2}\,,\label{eq:symmetry01}
\end{equation}
with $\tau\neq 0,\pi$ and $j$ is 1 or 2 or 3 (at will). For example, for the $j=3$ case, this explicitely gives
\begin{equation}
\Delta_1=\begin{pmatrix} \vartriangle & \vartriangle & 0 \\ \vartriangle & \vartriangle & 0 \\ 0 & 0 & 0 \end{pmatrix},\
\Delta_2=\begin{pmatrix} 0 & 0 & 0 \\ 0 & 0 & 0 \\ 0 & 0 & \blacktriangle \end{pmatrix},\label{eq:DeltaMat01}
\end{equation}
\begin{equation}
\Gamma_1=\begin{pmatrix} \triangledown & \triangledown & \triangledown \\ \triangledown & \triangledown & \triangledown \\ 0 & 0 & 0 \end{pmatrix},\
\Gamma_2=\begin{pmatrix} 0 & 0 & 0 \\ 0 & 0 & 0 \\ \blacktriangledown & \blacktriangledown & \blacktriangledown \end{pmatrix},\label{eq:GammaMat01}
\end{equation}
where $\vartriangle$, $\blacktriangle$, $\triangledown$ and $\blacktriangledown$ are the different generic entries allowed by \refeq{eq:symmetry01} to be non-zero. The corresponding mass matrices in \refeq{eq:MassMat:01} are
\begin{equation}
\Mu^0=\begin{pmatrix}v_1\begin{pmatrix}\vartriangle & \vartriangle \\ \vartriangle & \vartriangle \end{pmatrix} & \begin{matrix} 0\\ 0\end{matrix} \\  \quad\begin{matrix} 0& 0\end{matrix} & v_2e^{i\theta}(\blacktriangle)\end{pmatrix}
\end{equation}
and
\begin{equation}
\Md^0=\begin{pmatrix}v_1\begin{pmatrix} \triangledown & \triangledown & \triangledown \\ \triangledown & \triangledown & \triangledown \end{pmatrix}  \\  v_2e^{i\theta}\begin{pmatrix} \blacktriangledown & \blacktriangledown & \blacktriangledown \end{pmatrix}\end{pmatrix},
\end{equation}
while 
\begin{equation}
\Nu^0=\begin{pmatrix}v_2\begin{pmatrix}\vartriangle & \vartriangle \\ \vartriangle & \vartriangle \end{pmatrix} & \begin{matrix} 0\\ 0\end{matrix} \\  \quad\begin{matrix} 0& 0\end{matrix} & -v_1e^{i\theta}(\blacktriangle)\end{pmatrix}
\end{equation}
and 
\begin{equation}
\Nd^0=\begin{pmatrix}v_2\begin{pmatrix} \triangledown & \triangledown & \triangledown \\ \triangledown & \triangledown & \triangledown \end{pmatrix}  \\  -v_1e^{i\theta}\begin{pmatrix} \blacktriangledown & \blacktriangledown & \blacktriangledown \end{pmatrix}\end{pmatrix}.
\end{equation}
$\Mu^0$ and $\Nu^0$ are simultaneously diagonalised, 
\begin{equation}
\Nu=\frac{v_2}{v_1}\text{diag}(m_u,m_c,0)-\frac{v_1}{v_2}\text{diag}(0,0,m_t),
\end{equation}
and it is important to stress that $U_{uL}$ has the following block form
\begin{equation}
U_{uL}=\begin{pmatrix} \times & \times & 0\\ \times & \times & 0 \\ 0 & 0 & 1\end{pmatrix}.\label{eq:Uul01}
\end{equation}
On the other hand, $\Md^0$ and $\Nd^0$ are not simultaneously diagonalised since
\begin{multline}
\Nd^0=\frac{v_2}{v_1}\Md^0-\frac{v_2}{\sqrt 2}\left(\frac{v_2}{v_1}+\frac{v_1}{v_2}\right)e^{i\theta}\Gamma_2\Rightarrow \\
\Nd=\frac{v_2}{v_1}\Md-\frac{v_2}{\sqrt 2}\left(\frac{v_2}{v_1}+\frac{v_1}{v_2}\right)e^{i\theta}U^\dagger_{dL}\Gamma_2 U_{dR}.\label{eq:FCNC01}
\end{multline}
The last term in \refeq{eq:FCNC01} concentrates all the difficulties if we want it to be related to masses, mixings and $\tan\beta$: $U_{dR}$ appears explicitely and instead of the full mixing matrix $V=U^\dagger_{uL}U_{dL}$, just $U_{dL}$ is involved. One can understand how these difficulties are bypassed by \refeq{eq:symmetry01} in the following way: if $\Gamma_2\propto P\Md^0$ with $P$ some fixed matrix, $U_{dR}$ can be traded for $U_{dL}$ since
\[
\Gamma_2 U_{dR}\propto P\Md^0 U_{dR}=P U_{dL} \Md.
\]
With $\Gamma_2$ in \refeq{eq:GammaMat01}, $v_2e^{i\theta}\Gamma_2=\sqrt 2 P \Md^0$ where
\[
P=\begin{pmatrix} 0 & 0 & 0\\ 0 & 0 & 0\\ 0 & 0 & 1\end{pmatrix}.
\]
Then,
\[
[U^\dagger_{dL} P U_{dL}]_{ij}=[U^\ast_{dL}]_{3i}\,[U^{\phantom\ast}_{dL}]_{3j},
\]
but, because of the block form of $U_{uL}$ in \refeq{eq:Uul01}, the elements in the third row of the CKM matrix, $V_{3i}$, are simply $V_{3i}=[U_{dL}]_{3i}$ and thus 
\refeq{eq:FCNC01} becomes
\begin{equation}
\Nd=\frac{v_2}{v_1}[\Md]_{ij}-\left(\frac{v_2}{v_1}+\frac{v_1}{ v_2}\right){V}^\ast_{3i}{V}^{\phantom\ast}_{3j}[\Md]_{jj}
\end{equation}
where, as anticipated, FCNC appear at tree level and are controlled by fermion masses, CKM elements and $\tan\beta$. For this example, $\Nd$ reads 
\begin{multline*}
\Nd=\frac{v_2}{v_1}\begin{pmatrix}m_d&0&0\\ 0&m_s&0\\ 0&0&m_b\end{pmatrix}\\-\left(\frac{v_1}{v_2}+\frac{v_2}{v_1}\right)
\begin{pmatrix} 
m_d |V_{td}|^2 & m_s V_{td}^\ast V_{ts}^{\phantom{\ast}} & m_b V_{td}^\ast V_{tb}^{\phantom{\ast}}\\
m_d V_{ts}^\ast V_{td}^{\phantom{\ast}} & m_s |V_{ts}|^2 & m_b V_{ts}^\ast V_{tb}^{\phantom{\ast}}\\
m_d V_{tb}^\ast V_{td}^{\phantom{\ast}} & m_s V_{tb}^\ast V_{ts}^{\phantom{\ast}} & m_b |V_{tb}|^2
\end{pmatrix}.
\end{multline*}
This detailed example, starting in \refeq{eq:symmetry01}, is just one case: in (\ref{eq:symmetry01}) one can as well choose $j=1$ or $j=2$; furthermore, instead of $u_{Rj}^{0}\mapsto e^{i2\tau}u_{Rj}^{0}$ and $d_{Rj}^0\mapsto d_{Rj}^0$, choosing $d_{Rj}^{0}\mapsto e^{i2\tau}d_{Rj}^{0}$ and $u_{Rj}^0\mapsto u_{Rj}^0$, would lead to tree level FCNC in the \emph{up} sector instead, giving \emph{six} different cases in the quark sector. For the lepton sector, this same reasoning leads to six possibilities too\footnote{Furthermore, as already mentioned, addressing the lepton sector is necessary as soon as the renormalization group evolution of the Yukawa couplings is considered \cite{Botella:2011ne}.}, for an overall {\bf thirty-six} different models\footnote{For Majorana neutrinos, the implementation of the symmetry is restricted to models with tree level FCNC in the charged lepton sector, and the overall number is reduced to eighteen \cite{Botella:2011ne}.}. In the following, models are labelled by the right-handed fermions transforming non-trivially: e.g., the previous explicit example has $u_{Rj}^{0}\mapsto e^{i2\tau}u_{Rj}^{0}$ for $j=3$, the corresponding quark label would simply be $t$ (in addition a \emph{lepton} label is also required to fully define the model). For a detailed account of how BGL models can be understood in terms of a Minimal Flavour Violating expansion based on projection operators \cite{Botella:2004ks}, see \cite{Botella:2009pq}. Having presented the general properties of BGL models, we can now concentrate on the different processes which might be affected by non-SM contributions and could provide constraints on the new parameters of interest: the masses of the new scalars -- $m_R$, $m_I$ and $m_{H^\pm}$ -- and $\tan\beta$.

\section{Constraints}
\label{sec:Constraints}
Following \cite{Botella:2014ska} we organise flavour changing processes of interest in terms of the nature of the New Physics contributions that are involved:
\begin{itemize}
\item Processes with contributions mediated by $H^\pm$ at tree level in addition to the SM $W^\pm$-mediated tree level ones, such as universality in lepton decays or leptonic and semileptonic decays like $\pi\to e\nu$, $B\to\tau\nu$ or $\tau\to M\nu$.
\item Processes with contributions mediated by the neutral scalars $R$, $I$, at tree level and
\begin{itemize}
\item contributions from the SM at the loop level as in $B_{s,d}\to\mu^+\mu^-$ or neutral meson oscillations,
\item negligible SM loop contributions as in $\tau^-\to\mu^-\mu^-\mu^+$ or $\mu^-\to e^-e^-e^+$.
\end{itemize}
\item Processes with NP loop contributions and
\begin{itemize}
\item SM loop contributions as in $B\to X_s\gamma$,
\item negligible SM loop contributions as in $\tau\to\mu\gamma$ or $\mu\to e\gamma$.
\end{itemize}
\end{itemize}
In addition, electroweak precision contraints are incorporated through $Z\to b\bar b$ and oblique $S$ and $T$ constraints. Table \ref{TAB:sum} provides a summary of the constraints.  
\begin{table}[h] 
\begin{center}
{\small
\begin{tabular}{c|cc|cc|cc|}
\cline{2-7} & \multicolumn{4}{|c|}{BGL} & \multicolumn{2}{||c|}{SM}\\ 
\cline{2-7} & \multicolumn{2}{|c|}{$H^\pm$} & \multicolumn{2}{|c|}{$R$, $I$} & \multicolumn{1}{||c|}{\multirow{2}{*}{Tree}} & \multicolumn{1}{|c|}{\multirow{2}{*}{Loop}}\\
\cline{2-5} & \multicolumn{1}{|c|}{Tree} & \multicolumn{1}{|c|}{Loop} & \multicolumn{1}{|c|}{Tree} & \multicolumn{1}{|c|}{Loop} & \multicolumn{1}{||c|}{} & \multicolumn{1}{|c|}{} \\
\hline\multicolumn{1}{|c|}{$M\to\ell\bar\nu,M^\prime\ell\bar\nu$} & \multicolumn{1}{|c|}{\nchck} & \grchck & \multicolumn{1}{|c|}{} & \multicolumn{1}{|c|}{\grchck} & \multicolumn{1}{||c|}{\nchck} & \multicolumn{1}{|c|}{\grchck}\\
\hline\multicolumn{1}{|c|}{Universality} & \multicolumn{1}{|c|}{\nchck} & \grchck & \multicolumn{1}{|c|}{} & \multicolumn{1}{|c|}{\grchck} & \multicolumn{1}{||c|}{\nchck} & \multicolumn{1}{|c|}{\grchck}\\
\hline\hline\multicolumn{1}{|c|}{$M^0\to\ell_1^+\ell_2^-$} & \multicolumn{1}{|c|}{} & \grchck & \multicolumn{1}{|c|}{\nchck} & \multicolumn{1}{|c|}{\grchck} & \multicolumn{1}{||c|}{} & \multicolumn{1}{|c|}{\nchck}\\
\hline\multicolumn{1}{|c|}{$M^0\rightleftarrows \bar M^0$} & \multicolumn{1}{|c|}{} & \grchck & \multicolumn{1}{|c|}{\nchck} & \multicolumn{1}{|c|}{\grchck} & \multicolumn{1}{||c|}{} & \multicolumn{1}{|c|}{\nchck}\\
\hline\multicolumn{1}{|c|}{$\ell_1^-\to\ell_2^-\ell_3^+\ell_4^-$} & \multicolumn{1}{|c|}{} & \grchck & \multicolumn{1}{|c|}{\nchck} & \multicolumn{1}{|c|}{\grchck} & \multicolumn{1}{||c|}{} & \multicolumn{1}{|c|}{\grchck}\\
\hline\hline\multicolumn{1}{|c|}{$B\to X_{s}\gamma$} & \multicolumn{1}{|c|}{} & \nchck & \multicolumn{1}{|c|}{} & \multicolumn{1}{|c|}{\nchck} & \multicolumn{1}{||c|}{} & \multicolumn{1}{|c|}{\nchck}\\
\hline\multicolumn{1}{|c|}{$\ell_j\to \ell_i\gamma$}  & \multicolumn{1}{|c|}{} & \nchck  & \multicolumn{1}{|c|}{} & \multicolumn{1}{|c|}{\nchck} & \multicolumn{1}{||c|}{} & \multicolumn{1}{|c|}{\grchck}\\
\hline\hline\multicolumn{1}{|c|}{EW Precision} & \multicolumn{1}{|c|}{} & \nchck & \multicolumn{1}{|c|}{} & \multicolumn{1}{|c|}{\nchck} & \multicolumn{1}{||c|}{} & \multicolumn{1}{|c|}{\nchck}\\
\hline
\end{tabular}
} 
\end{center}
\caption{Summary table from \cite{Botella:2014ska}; leading contributions are tagged $\nchck$ while subleading or negligible ones are tagged $\grchck$.\label{TAB:sum}}
\end{table}

For a detailed account of the processes and the numerical values used as input in the analysis, see section 3 and appendices A and B of reference \cite{Botella:2014ska}.

\section{Results}
\label{sec:Results}
\begin{figure}[t]
\centering
\includegraphics[width=0.45\textwidth]{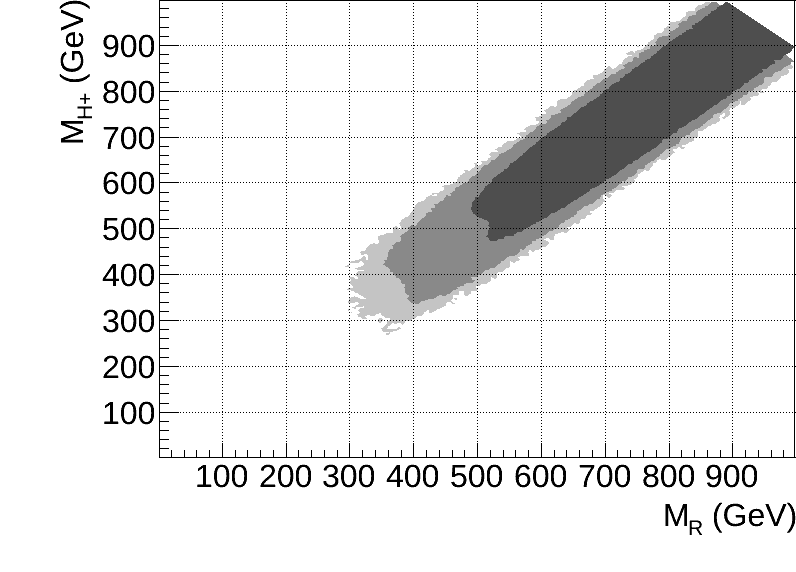}\\
\includegraphics[width=0.45\textwidth]{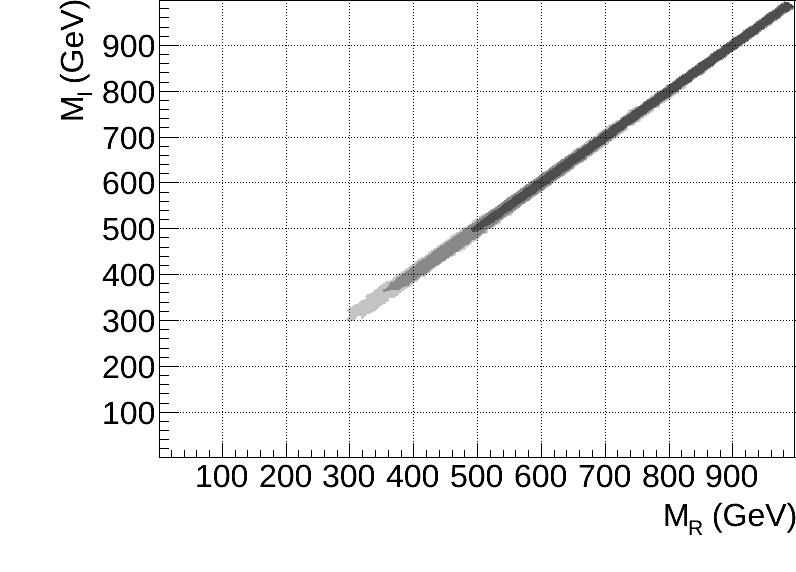}
\caption{Scalar masses allowed by the constraints from the oblique parameters in model $(t,\tau)$.}
\label{fig:ScalarMasses}
\end{figure}

\begin{sidewaysfigure}
\begin{minipage}{0.45\textheight}
\vspace{9cm}
\begin{center}
\includegraphics[width=0.95\textwidth]{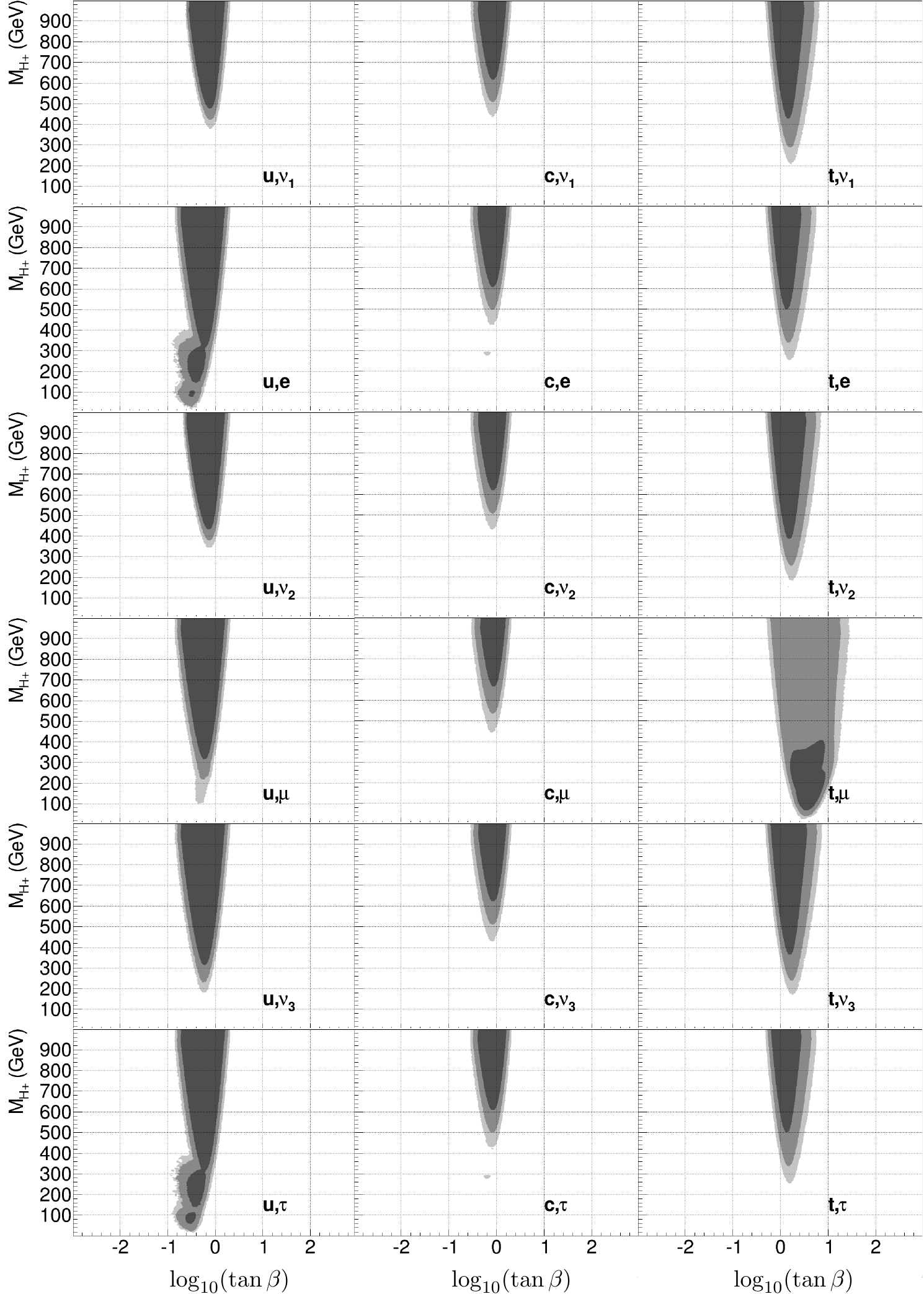}
\caption{Up models\label{fig:all1}}
\end{center}
\end{minipage}\hspace{1cm}
\begin{minipage}{0.45\textheight}
\vspace{9cm}
\begin{center}
\includegraphics[width=0.95\textwidth]{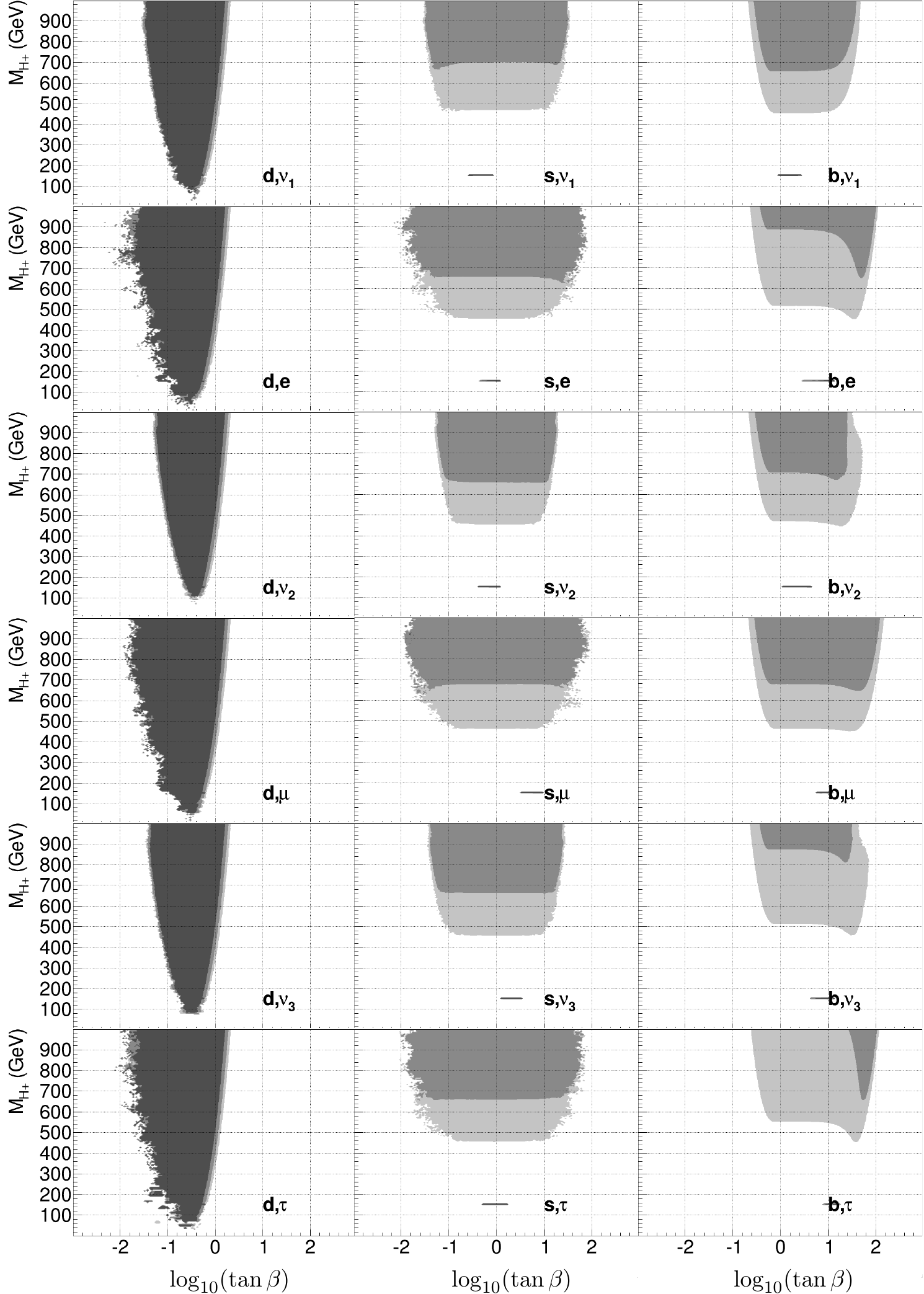}
\caption{Down models\label{fig:all2}}
\end{center}
\end{minipage}
\begin{center}
Summary plots of the allowed 68\%, 95\% and 99\% CL regions in $m_{H^\pm}$ vs. $\log_{10}\tan\beta$ for all BGL models.
\end{center}\label{fig:all}
\end{sidewaysfigure}

The main goal of this analysis is understanding which are the values of the new parameters -- $v_2/v_1=\tan\beta$ and masses of the scalars $m_I$, $m_R$ and $m_{H^\pm}$ -- allowed when all the previous constraints are imposed. This task is simplified by first paying attention to the effect of the precision electroweak constraints, in particular the oblique parameters. For similar values of $m_I$, $m_R$ and $m_{H^\pm}$, the oblique parameters are in good agreement with experimental data \cite{Botella:2014ska}. This fact is illustrated in figure \ref{fig:ScalarMasses}, where the allowed regions corresponding to the usual 68\%, 95\% and 99\% confidence levels are shown for a particular model, $(t,\tau)$, after the constraints on the oblique parameters are imposed. Therefore, although all three masses are varied independently in the analyses, only results for one of them, $m_{H^\pm}$, are displayed. 
Figures \ref{fig:all1} and \ref{fig:all2}, taken from \cite{Botella:2014ska}, are the central result: they collect the allowed regions (68\%, 95\% and 99\% confidence levels) in terms of $m_{H^\pm}$ and $\tan\beta$ for all 36 models. Some comments are in order.
\begin{itemize}
\item Although tree level FCNC experimental constraints on down quark models are a priori tighter than on up quark models, up quark models are not less constrained due to the impact of $b\to s\gamma$ on the allowed $H^\pm$ masses.
\item Since $t$ and $b$ models give a stronger FCNC suppression -- because of the hierarchical structure of the CKM matrix --, one would expect them to be less constrained; nevertheless, the effect of $b\to s\gamma$ partially changes that picture and $d$ models are indeed less constrained than $b$ ones.
\item It is important to stress that because of the strong suppression achieved for tree level FCNC, the importance of $H^\pm$ as a window to New Physics is enhanced, as the significant role played by $b\to s\gamma$ shows.
\item Notice however that, although $b\to s\gamma$ provides relevant constraints, it is not as determinant as in type II 2HDM where it automatically forces $m_{H^\pm}>380$ GeV \cite{Hermann:2012fc}; this is due to the different $\tan\beta$ dependence in BGL models with respect to type II 2HDM (in addition, this dependence changes in the different BGL models).
\item Concerning the leptonic part, since tree level neutrino FCNC are irrelevant (because of the small neutrino masses), $e$, $\mu$ and $\tau$ are a priori less contrained than their neutrino counterparts, nevertheless such differences are insignificant: leptonic constraints are thus secondary once the effect of other constraints is considered.
\item Lower bounds on the scalar masses are in the $100-400$ GeV ballpark for many models, opening the window to potential direct searches at the LHC. There are, however, some exceptions: models of types $s$ and $b$ require masses above $500-600$ GeV (although a wider range of $\tan\beta$ values is acceptable in those models).
\item Finally, it has to be stressed that models of type $t$, the Minimal Flavour Violating ones in the sense of \cite{Buras:2000dm,D'Ambrosio:2002ex}, are promising but not unique: other models allow for light scalars.
\end{itemize}

\section{Conclusions}
\label{sec:Conclusions}
Two Higgs doublet models of the Branco-Grimus-Lavoura class are a viable scenario where tree level Flavour Changing Neutral Currents arise in a controlled manner: they are proportional to mixings, fermion masses and $\tan\beta$. The present study shows that despite the existing tight experimental constraints, several types of BGL models are of immediate interest since they can accommodate new scalars light enough to be within direct experimental reach of the LHC.

\section*{Acknowledgments}
\label{sec:Acknowledgments}
\noindent MN thanks F.J. Botella, G.C. Branco, M.N. Rebelo and L. Pedro for constructive conversations and comments, and acknowledges financial support from \emph{Funda\c c\~ao para a Ci\^encia e a Tecnologia} (FCT, Portugal) through projects CFTP-FCT Unit 777 and CERN/FP/123580/2011, and MINECO (Spain) through grant FPA2011-23596.

\clearpage

%
%
%




\clearpage








\end{document}